\documentclass{article}

\usepackage[preprint]{neurips_2021}

\usepackage[utf8]{inputenc} 
\usepackage[T1]{fontenc}    
\usepackage{hyperref}       
\usepackage{url}            
\usepackage{booktabs}       
\usepackage{amsfonts}       
\usepackage{nicefrac}       
\usepackage{microtype}      
\usepackage{xcolor}         
\usepackage{multirow}
\usepackage{graphicx}
\usepackage{amsmath}
\usepackage[linesnumbered,ruled]{algorithm2e}
\usepackage{bm}
\title{Simultaneous boundary shape estimation and velocity field de-noising in Magnetic Resonance Velocimetry using Physics-informed Neural Networks}

\author{%
  Ushnish Sengupta \\
  Department of Engineering\\
  University of Cambridge\\
  Cambridge, CB2 1PZ \\
  \texttt{us271@cam.ac.uk} \\
  \And
  Alexandros Kontogiannis\\
  Department of Engineering\\
  University of Cambridge\\
  Cambridge, CB2 1PZ \\
  \texttt{ak2239@cam.ac.uk} \\
  \AND
  Matthew P. Juniper\\
  Department of Engineering\\
  University of Cambridge\\
  Cambridge, CB2 1PZ \\
  \texttt{mpj1001@cam.ac.uk} \\
}

\begin{document}

\maketitle

\begin{abstract}
Magnetic resonance velocimetry (MRV) is a non-invasive experimental technique widely used in medicine and engineering to measure the velocity field of a fluid. These measurements are dense but have a low signal-to-noise ratio (SNR). The measurements can be de-noised by imposing physical constraints on the flow, which are encapsulated in governing equations for mass and momentum. Previous studies have required the shape of the boundary (for example, a blood vessel) to be known \textit{a priori}. This, however, requires a set of additional measurements, which can be expensive to obtain. In this paper, we present a physics-informed neural network that instead uses the noisy MRV data alone to simultaneously infer the most likely boundary shape and de-noised velocity field. We achieve this by training an auxiliary neural network that takes the value 1.0 within the inferred domain of the governing PDE and 0.0 outside. This network is used to weight the PDE residual term in the loss function accordingly and implicitly learns the geometry of the system. We test our algorithm by assimilating both synthetic and real MRV measurements for flows that can be well modeled by the Poisson and Stokes equations. We find that we are able to reconstruct very noisy (SNR = 2.5) MRV signals and recover the ground truth with low reconstruction errors of $3.7 - 7.5 \%$. The simplicity and flexibility of our physics-informed neural network approach can readily scale to assimilating MRV data with complex 3D geometries, time-varying 4D data, or unknown parameters in the physical model. 
\end{abstract}

\section{Introduction}
\subsection{Magnetic resonance velocimetry: promises and challenges}

Magnetic resonance imaging (MRI) and magnetic resonance velocimetry (MRV) are non-invasive imaging techniques that are used in health assessment and numerous applications in engineering \citep{Fukushima1999,VanDeMeent2010,Blythe2015,Gladden2017}. Magnetic resonance velocimetry, in particular, is widely used in order to investigate anomalies in the cardiovascular system, such as aneurysms and blockages in the blood vessels (stenoses). It provides a non-invasive way to obtain volumetric measurements of time-varying 3D velocity fields in complicated geometries \citep{Markl2012}, and is usually preferred over X-ray computed tomography scans because it does not involve ionizing radiation. However, the long acquisition time due to repeated scans hinders its use in many cases. To circumvent this problem, fast acquisition protocols (pulse sequences) can be used, although this usually produces images with artefacts and low signal-to-noise ratio (SNR). Another way to accelerate signal acquisition is by using sparse sampling patterns and subsequently reconstructing the signal. The latter approach is commonly referred to as \textit{compressed sensing} \citep{Donoho2006,Lustig2007,Benning2014}, where \textit{a priori} knowledge about the structure of the data is injected by considering a general regularization norm (e.g. total variation, wavelets as filter), but without considering the physics of the problem. General mathematical methods for image processing have also been used to reconstruct and segment objects from MRV images \citep{Debroux2019,Corona2019} and, more recently, deep variational neural networks have shown promising results in reconstructing MRI and MRV images \citep{Hammernik2018,Vishnevskiy2020}. However, the above reconstruction/segmentation methods fail to take into account the physical knowledge of the systems they are imaging, which can often be described by a PDE.


\subsection{Using physics to gain resolution}

Prior work on fast acquisition protocols and compressed sensing has relied on ad-hoc principles to reconstruct MRV velocity fields, while machine learning approaches make use of libraries of patient data for training, which can be hard to obtain and introduce brittleness into the system. There is, however, a promising alternative way to reconstruct noisy flow images that makes use of the equations of fluid dynamics. One can formulate an inverse flow problem, where certain model parameters are updated in an iterative way so that the modeled flow matches the measured flow. This not only provides noiseless velocity fields, but it can also be used to infer the unknown model parameters; parameters which the MRV experiment cannot measure (e.g. pressure and viscosity). Subsequently, this implies that the inferred model parameters can be used to simulate different flow conditions for a specific patient (patient-specific modeling) \citep{Taylor2009,Morris2016}. 

Recent work using purely PDE (adjoint-based) methods involves the inverse problem of finding the inlet velocity boundary condition of a Navier--Stokes problem, in order to reconstruct time-varying 3D velocity fields in blood vessel replicas \citep{Koltukluoglu2018,Koltukluoglu2019,Koltukluoglu2019b,Funke2019}. This approach usually relies on the theoretical derivation of an Euler--Lagrange system, the design of an algorithm to decouple the system (segregated approach), and the numerical solution of the individual PDEs. Around the same time, physics-informed neural networks (PINNs) were introduced as a tool that can integrate data and physics to solve general inverse problems by leveraging the function approximation capabilities of neural networks \citep{Raissi2019,Sun2020,Doan2020}. \cite{Raissi2020} used PINNs to reconstruct fluid flows and infer unknown flow quantities. Despite technical differences, both approaches share the same principle: solve an inverse problem by leveraging physical knowledge in the form of a PDE.
  
\subsection{Our contribution}

In this paper we propose a physics-informed neural network that simultaneously \textit{reconstructs and segments} noisy MRV images. Previous studies have assumed that the geometry of the system is known \textit{a priori}. In practice, computed tomography (CT) or magnetic resonance angiography (MRA) scans are often needed in order to find the geometry of the blood vessel. This geometry subsequently has to be reconstructed, segmented, and smoothed. This process requires an additional experiments, substantial effort, and can introduce geometric uncertainties and mismatches between the flow domain and the actual geometry \citep{Katritsis2007,Sankaran2016}. Therefore, we propose a consistent approach to both problems of inferring the boundary shape and filtering the velocity fields using only MRV signals. Our approach uses a PINN augmented with an auxiliary neural network which implicitly infers the shape of the domain and weights the PINN loss function accordingly.

The notebooks and the datasets used to produce the results in this paper are hosted at \url{https://anonymous.4open.science/r/MRV-Denoising-3FC2/}.

\section{Methods}
\label{gen_inst}

\subsection{Physics}
As a proof of concept, we investigate two kinds of simple, two-dimensional, steady flows in this paper. The first is the fully developed laminar flow of a Newtonian fluid through a cross-section. This is governed by the well-known Poisson's equation with Dirichlet conditions $v = 0$ at the boundary.

\begin{equation}
    \Delta v = f
\end{equation}

where $f = -\frac{1}{\eta}\frac{\partial P}{\partial z}$, $\eta$ being the dynamic viscosity and $\frac{\partial P}{\partial z}$ the pressure gradient. The PDE residual $\mathcal{R}(\bold{x})$ that needs to be entered into the PINN loss function in this case is simply $\mathcal{R}(\bold{x}) = \Delta v - f$.

We also consider 2D steady creeping flow which is governed by the Stokes equations, a special case of the Navier--Stokes equations which arises if the flow Reynolds number $Re \to 0{}$. 
\begin{equation}
    \nabla \cdot {\bold{v}} = 0
\end{equation}

\begin{equation}    
    \mu \nabla ^2 \bold{v} - \nabla \cdot {P}= 0
\end{equation}

In their original form, the Stokes equations are inconvenient because the unknown pressure field $P$, for which no data is available, would need to be learned. We circumvent this by using the biharmonic streamfunction form of the 2D Stokes equations in our PINN, which eliminates the pressure field.

\begin{equation}    
    \Delta^2 \cdot {\psi} = 0
\end{equation}

The velocity field is given by $\bold{v} = \nabla \times \psi$ and the residual is simply $\mathcal{R}(\bold{x}) = \Delta^2 \bold{\psi}$.
To assimilate flows governed by the Navier--Stokes equations, one would similarly use the vorticity transport formulation, which eliminates the pressure field from the momentum equation by taking the curl.
\subsection{Neural network architecture and loss function design}

\begin{figure}
\centering
\includegraphics[scale = 0.475]{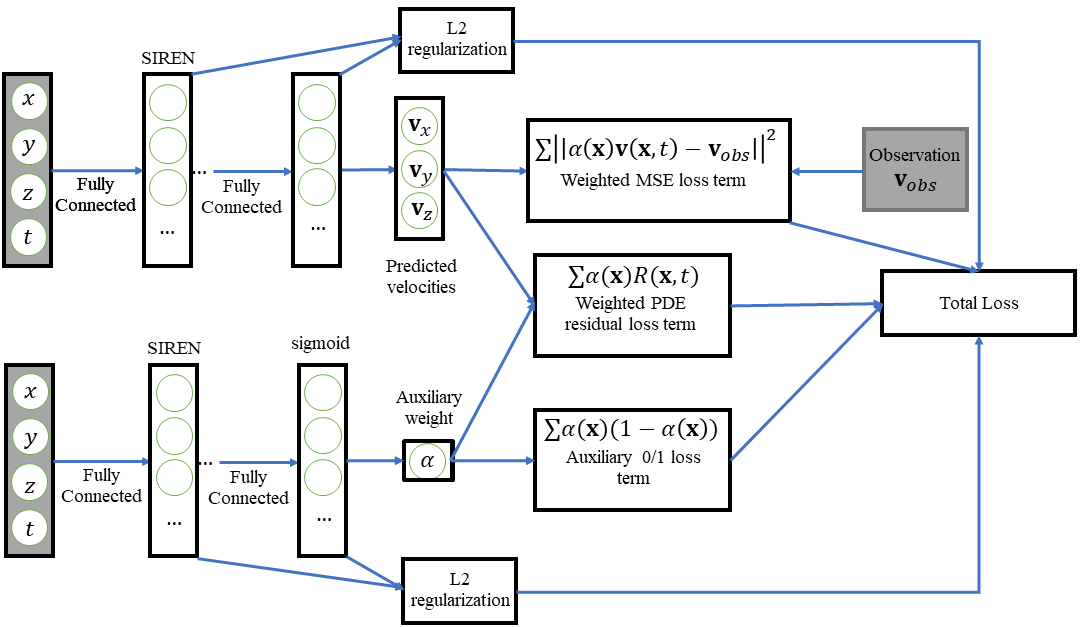}
\caption{Architecture of the MRV-denoising PINN augmented with auxiliary network for boundary shape inference.}
\label{NN_arch}
\end{figure}

We use two neural networks: a solution network $\bold{v}_{sol}(\bold{x}; \theta_{sol})$ which learns a denoised velocity field that respects both the data and the PDE inside the domain, and an auxiliary NN $\alpha(\bold{x}; \theta_{aux})$ which infers the shape of the domain. The auxiliary network is allowed to take the values of either 0 or 1 within the bounding box, which serves as an indicator of whether a point is outside the domain or inside, respectively. The two networks have a simple MLP architecture with fully connected layers. The hidden layers use periodic SIREN activation functions \citep{sitzmann2020implicit}, which have been shown to dramatically improve convergence and approximation quality when using neural networks to approximate PDE solutions, compared with other differentiable activations. The final output of the auxiliary network uses a sigmoid activation to restrict its value between 0 and 1.

The loss function of our augmented PINN has 4 terms. The first is the mean squared error term $\frac{1}{n_d}\sum \| \bold{v}_{mini} - \alpha( \bold{x}_{mini}; \theta_{aux}) \bold{v}_{sol}(\bold{x}_{mini}; \theta_{sol}) \|^2 $. Within the domain, where $\alpha$ should be 1, this reduces to the ordinary mean squared error and outside the domain, where $\alpha = 0$, it is still consistent because $\bold{v}_{true} = 0$ there. The second term contains the squared sum of PDE residuals $\frac{w_{res}}{n_r} \sum \alpha( \bold{x}_{d} ; \theta_{aux}) \mathcal{R}(\bold{x}_{d}; \theta_{sol}, p)^2$ evaluated at randomly sampled points $\bold{x}_{d}$, which are similarly weighted using $\alpha$ because we only require the PDE to be respected inside the domain. The third term $\frac{1}{n_r} \sum \alpha( \bold{x}_{d}; \theta_{aux}) (1 - \alpha( \bold{x}_{d}; \theta_{aux}))$ keeps the auxiliary function from assuming values other than 0 and 1 within the bounding box. The final term is simply an L2 regularization for the two networks. The weight of the L2 regularization term $w_{reg}$ is important for the auxiliary network, because it dictates the smoothness of the boundary that will be inferred.

It is crucial to train $v_{sol}(\bold{x}; \theta_{sol})$ and $\alpha(\bold{x}; \theta_{aux})$ using only the observed data in the beginning. If the loss function above is optimized without a reasonable initial guess, the algorithm often fails to converge to the true solution. 

\begin{algorithm}
    \SetKwInOut{Input}{Input}
    \SetKwInOut{Output}{Output}
    
    \Input{ Noisy velocity observations $\{\bold{x}_{obs}, \bold{v}_{obs}\}$. Hyperparameters $N_l$, $N_n$, $v_{threshold}$, $n_{epochs}$, $lr$, $n_r$, $n_d$, $w_{res}$, $w_{reg}$.}
    \Output{ Denoised velocity field $\bold{v}_{sol}(\bold{x})$, inferred shape of the boundary, optional: unknown physical model parameter $p$.}
    
    Initialize the parameters of the solution neural network $\bold{v}_{sol}(\bold{x}; \theta_{sol})$ and auxiliary neural network $\alpha(\bold{x}; \theta_{aux})$ from a uniform distribution. $\theta_i \sim \mathcal{U}(-\sigma_i, \sigma_i)$, where $\sigma_i = 6.0/n_{in}$, $n_{in}$ being the number of inputs to the layer $\theta_i$ belongs to. The NNs have $N_l$ layers with $N_n$ nodes each and SIREN activations in the hidden layers. The auxiliary network must have a sigmoid activation in its final layer.
    
    An initial guess for the solution network $\bold{v}_{sol}(\bold{x}; \theta_{sol})$ is trained using the velocity observations $\{\bold{x}_{obs}, \bold{v}_{obs}\}$. L2 regularization is applied.
    
    An initial guess for the auxiliary neural network is then produced by training it on $\{\bold{x}_{obs}, H(\|\bold{v}_{obs}\| - v_{threshold})\}$, where $H$ is the Heaviside step function. L2 regularization is applied.
    
    \For{$j\gets1$ \KwTo $n_{epochs}$}{
    
        Sample $n_r$ points $\{\bold{x}_{d}\}$ uniformly within bounding box, sample minibatch of $n_d$ datapoints $\{\bold{x}_{mini}, \bold{v}_{mini}\}$.
        
        Compute PDE residuals of $\bold{v}_{sol}$ at sampled points $\mathcal{R}(\bold{x}_{d}; \theta_{sol}, p)$.
        
        Optimize loss $\mathcal{L} = \frac{1}{n_d}\sum \| \bold{v}_{mini} - \alpha( \bold{x}_{mini}; \theta_{aux}) \bold{v}_{sol}(\bold{x}_{mini}; \theta_{sol}) \|^2 + \frac{w_{res}}{n_r} \sum \alpha( \bold{x}_{d} ; \theta_{aux}) \mathcal{R}(\bold{x}_{d}; \theta_{sol}, p)^2 + \frac{1}{n_r} \sum \alpha( \bold{x}_{d}; \theta_{aux}) (1 - \alpha( \bold{x}_{d}; \theta_{aux})) + w_{reg}(\sum \|\theta_{sol,i}/\sigma_{sol,i}\|^2 + \sum \|\theta_{aux,i}/\sigma_{aux,i}\|^2)$ using ADAM with a learning rate $lr$ w.r.t. unknown parameters $\theta_{sol}, \theta_{aux}, p$.

    }
    
    $\bold{v}_{sol}(\bold{x}_k): = \bold{v}_{sol}(\bold{x}_k) H(\alpha(\bold{x}_k) - 0.5)$. Make the value of the velocity field zero where the auxiliary function is less than 0.5.
    
    Return $\bold{v}_{sol}(\bold{x})$, $p$ (optional), boundary shape from $\alpha = 0.5$ contour line.
    \caption{Denoising noisy MRV data and boundary shape inference with PINNs}
\end{algorithm}

We used the automatic differentiation library JAX \citep{jax2018github} to implement the augmented PINNs. All experiments were performed on a laptop with a 6 core, 2.60 GHz Intel i7-9750 CPU and NVIDIA RTX 3070 GPU. In the test cases we considered in Section \ref{exp}, the PINN takes $\sim 15-30$ minutes to converge to a reasonable solution.

We used number of layers $N_l = 6$ and number of nodes per layer $N_n = 32$ for all the test cases. These values were not chosen based on extensive tuning but rather this was found to be the smallest neural network that reliably and rapidly converged to a good solution. A $v_{threshold} = 15\%$ of the maximum velocity magnitude and  $n_r = n_d = 512 $ worked well for all the problems. The neural networks were trained until convergence and $n_{epochs}$ differed between problems, ranging from 40000 to 70000. Optimal values of $lr$, $w_{res}$ and $w_{reg}$ were obtained by tuning on noisy synthetic data of Poiseuille flow through a circular cross-section and Stokes flow through a straight channel. A general finding is that the PINN hyperparameters are fairly portable across problems which have the same governing equation, similar flow regimes and SNR. Convergence to the true solution is robust to small changes to hyperparameters.

\section{Experiments}
\label{exp}

\subsection{Synthetic data: Poisson}
We now test the algorithm for the problem of Poiseuille flow through a starfish-shaped pipe. The geometry of the starfish shape is given by $R(\theta) = 0.7 - 0.175\text{ cos}(5\theta) - 0.05\text{ cos}(6\theta) + 0.03\text{ sin}(9\theta)$. As discussed in section 2, this corresponds to a Poisson problem with zero Dirichlet boundary conditions on the wall. $f$ is assumed to be known and equal to -1.0. We create a synthetic MRV image by solving the Poisson problem using a PDE solver to obtain the ground truth velocity field $\bold{v}_{true}$ and then add Gaussian white noise with variance $\sigma^2$ to corrupt the image. The SNR for the corrupted image is 2.5.

We define the $L^2$ norm reconstruction error $\epsilon$ as follows:
\begin{equation}
    \epsilon = \frac{\|\bold{v}_{sol} - \bold{v}_{true}\|_{L^2}}{\|\bold{v}_{true}\|_{L^2}}
\end{equation}

Figure 2 shows the results of the denoising and shape inference algorithm. Hyperparameter values $lr = 3\times10^{-5}$, $w_{reg} = 10^{-8}$ and $w_{res} = 2.5\times 10^{-4}$ were used. The algorithm was run for 45000 epochs. We find that signal has been reconstructed accurately and the starfish shape of the domain has also been recovered. The reconstruction error $\epsilon$ is $\sim 4.4 \%$.

\begin{figure}
\centering
\includegraphics[scale = 0.25]{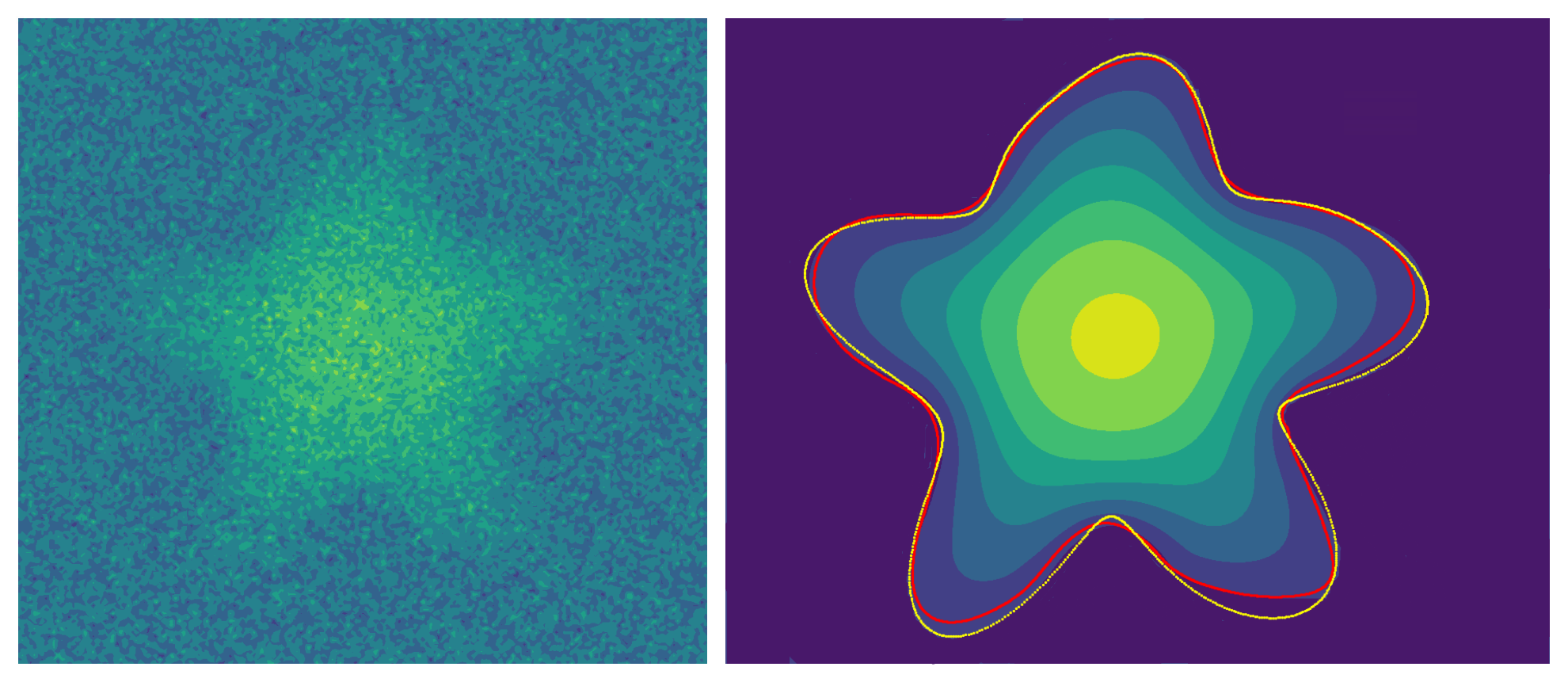}
\caption{Denoising and boundary inference for simulated Poiseuille flow through a starfish-shaped pipe cross-section. Original noisy signal to the left, denoised velocity field to the right. The ground truth boundary is shown in yellow, the inferred boundary is shown in red.}
\label{starfish}
\end{figure}

\subsection{Synthetic data: Stokes}
Next, we test the algorithm for 2D steady Stokes flow (figures 3, 4) in two different geometries: i) a curved channel and ii) a blood vessel dummy. As before, we add Gaussian white noise to the true solutions such that SNR = 2.5. The same hyperparameter values $lr = 10^{-5}$, $w_{reg} = 10^{-7}$ and $w_{res} = 10^{-6}$ were used in both cases. 

For the curved channel, we obtain an excellent reconstruction with $\epsilon = 3.7\%$ after running the algorithm for 50000 epochs. For the dummy blood vessel which is a more challenging test case, the results are slightly worse: $\epsilon = 7.5\%$ even after running the optimization for 70000 epochs. The geometry reconstruction is noticeably poor in regions of high curvature. In these regions, small shape changes do not affect the flow field strongly and as a result, the geometry is difficult to recover precisely from such a noisy signal. 
\begin{figure}
\centering
\includegraphics[scale = 0.175]{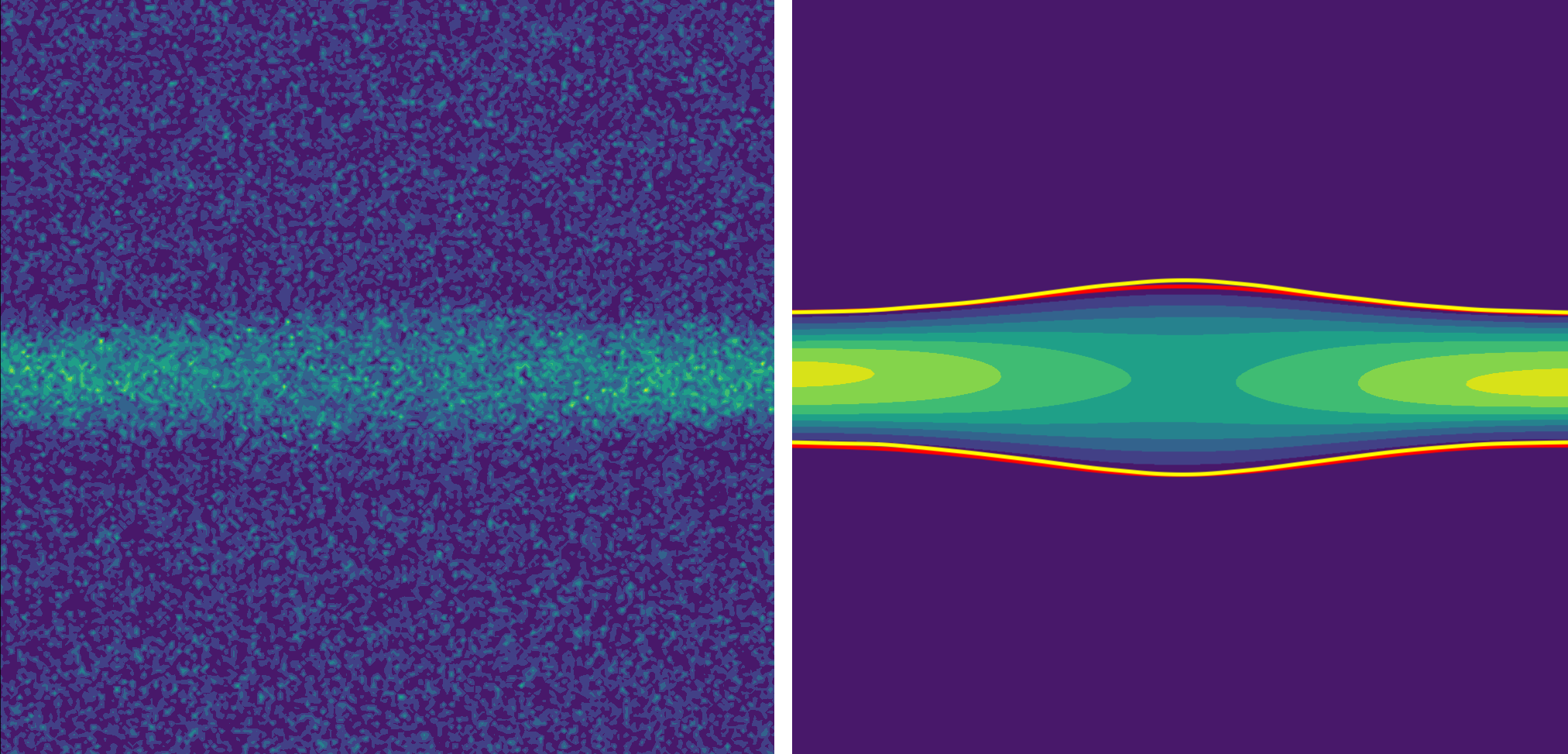}
\caption{Contour plot of velocity magnitude showing denoising and boundary inference for simulated Stokes flow data through a curved channel. Original noisy signal to the left, denoised velocity field to the right. The ground truth boundary is shown in yellow, the inferred boundary is shown in red.}
\label{curved}
\end{figure}

\begin{figure}
\centering
\includegraphics[scale = 0.16]{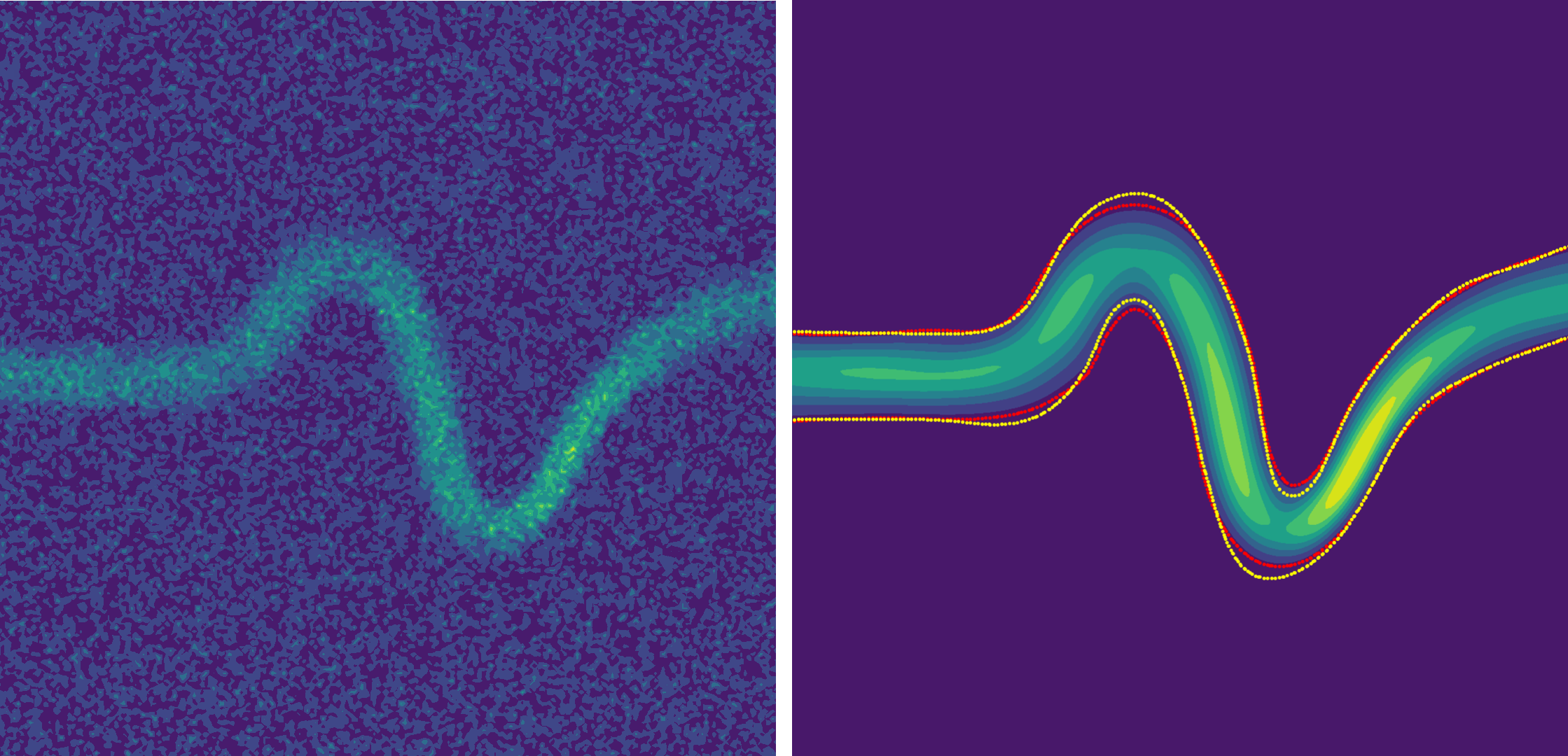}
\caption{Contour plot of velocity magnitude showing denoising and boundary inference for simulated Stokes flow through a dummy blood vessel. Original noisy signal to the left, denoised velocity field to the right. The ground truth boundary is shown in yellow, the inferred boundary is shown in red.}
\label{blood}
\end{figure}
\subsection{Real MRV data}
\begin{figure}
\centering
\includegraphics[scale = 0.155]{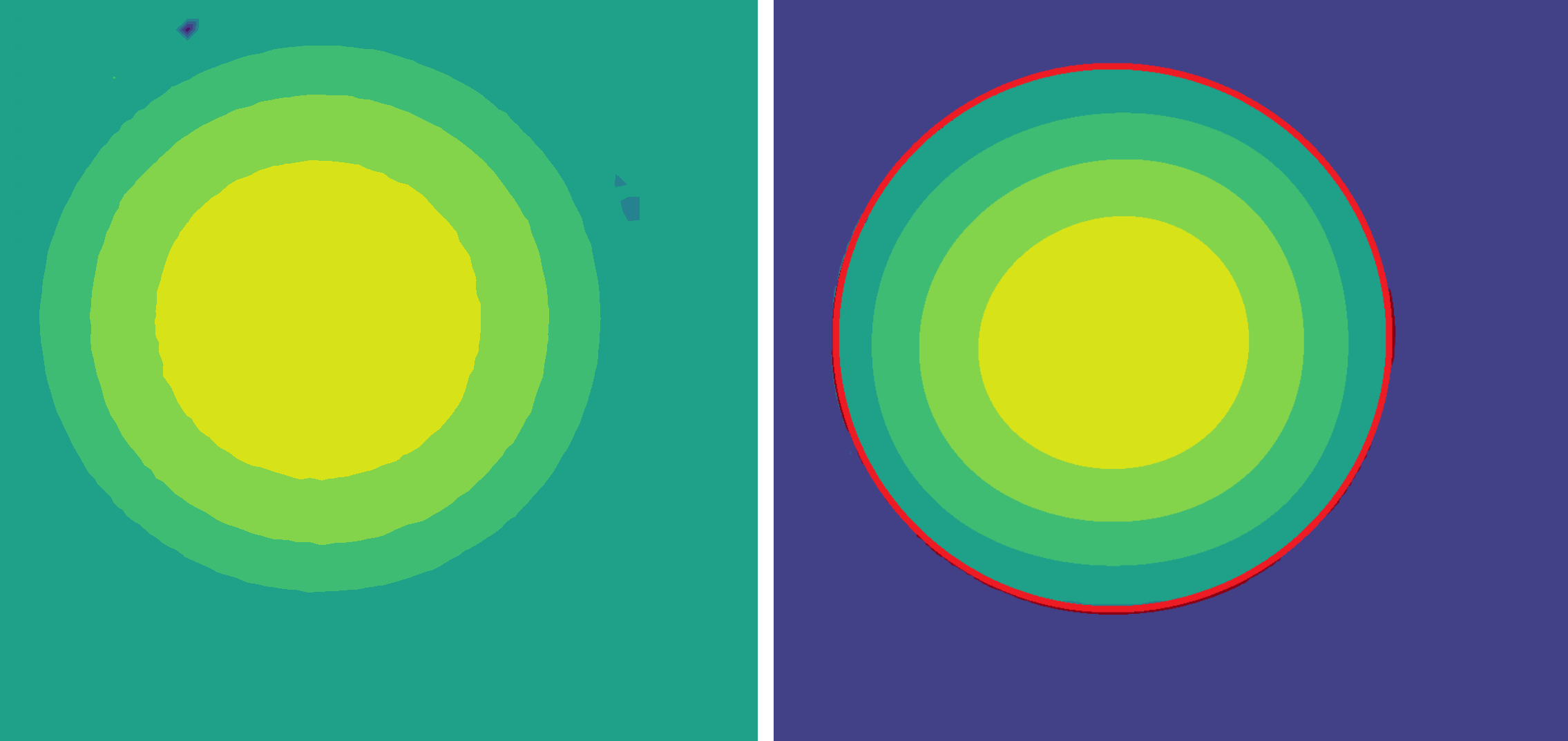}
\caption{Contour plot of velocity magnitude showing denoising and boundary inference for a real MRV image of nearly fully developed flow through a pipe. Original noisy MRV data to the left, denoised velocity field to the right. The inferred boundary is shown in red.}
\label{real}
\end{figure}

In this case, we consider a real MRV image ($64^2$ pixels) of almost fully developed laminar flow \citep{Reci2019} through a pipe with a circular cross-section. Unlike our synthetic problems, this data has very high SNR and aside from two small artifacts outside the domain, is unchallenging to assimilate. This case is still interesting, however, because the $f = -\frac{1}{\eta}\frac{\partial P}{\partial z}$ term in Poisson's equation is unknown to us and must be estimated during the optimization.

We use the same hyperparameters for this test case as the starfish problem. Figure \ref{real} shows the results. $f$ converged to the value $-0.0873$ from an initial guess of $-0.1$ which was estimated based on initial guesses for $\bold{v}_{sol}$ and $\alpha$. The regularized auxiliary neural network had no trouble ignoring the pixel artifacts.

\section{Conclusions}

We extend PINNs to solve inverse flow problems with noisy velocity observations when the shape of the boundary is not known.  This is achieved by introducing an auxiliary neural network with a sigmoid output which is allowed to take the values 1.0 or 0.0 and infers whether a point in the bounding box of the signal is within the domain or outside. As a proof-of-concept, we test our algorithm for the 2D Poisson and the Stokes problems using both real MRV measurements and noisy synthetic MRV measurements where the ground truth is known. We find that the algorithm is able to accurately reconstruct the ground truth signal with low errors ($\epsilon = 3.7-7.5\%$) and infer the shape of the boundary. For the real MRV data, we find that the algorithm is also able to estimate an unknown parameter in the physical model. Our approach has the potential to reconstruct very noisy MRV images and recover denoised flow fields that respect physical constraints, without the need for additional experiments to learn the domain geometry. 

One limitation of the current technique, which is shared by all PINN-based methods, is the relatively slow convergence. However, not putting a patient through a longer test sequence may be worth tens of minutes for post-processing. Additionally, although one of our goals is the eventual assimilation of time-varying data, this often involves moving and compliant boundaries, especially in cardiovascular medicine. In this case, it is not just fluid dynamics equations that need to be obeyed but the more complex equations of fluid-structure interaction. It is not immediately apparent how this should be handled within the current framework.

There are several exciting directions for future work. We are currently exploring flows that are governed by the Navier-Stokes equations instead of Stokes. This involves making a few additions to the residual term in the loss but otherwise, no other modification to the algorithm is necessary. The flexible nature of neural networks means that the this algorithm readily extends to 3D or 4D time-varying data with complex geometries, so assimilating realistic medical imaging data with this tool is a natural next step. As we move beyond the prototyping phase, we also need to focus more on optimizing our training times, using adaptive sampling approaches for PINNs or quantized inference algorithms. The capability of solving inverse problems with noisy data and unknown boundary geometry or parameters may have other applications beyond MRV signal filtering and these possibilities should also be explored.
\begin{ack}


\end{ack}

\bibliography{neurips}



\end{document}